\documentclass[twocolumn,showpacs,preprintnumbers,amsmath,amssymb,pra]{revtex4}

\usepackage{graphicx}
\usepackage{dcolumn}
\usepackage{color}
\usepackage{bm}
\usepackage{longtable}
\begin{document}

\title {\bf Distinguishability of apparatus states in quantum 
measurement in the Stern-Gerlach experiment}

\author{Dipankar Home
\footnote {dhome@bosemain.boseinst.ac.in}$^1$,
Alok Kumar Pan\footnote{apan@bosemain.boseinst.ac.in}$^1$, Md. 
Manirul Ali\footnote{manirul@imsc.res.in}$^2$ and A. S. Majumdar
\footnote{archan@bose.res.in}$^3$}             

\address{$^1$ Department of Physics, Bose Institute, Calcutta-700009, 
India}    

\address{$^2$ Institute of Mathematical Sciences, C.I.T. Campus, Taramani,
Chennai 600113, India}
                   
\address{$^3$ S. N. Bose National Centre for Basic Sciences, Block JD,
Sector III, Salt Lake, Calcutta-700098, India}
                                                               
\date{\today}

\begin{abstract}
In the context of the quantum mechanical modelling of a measurement process
using the Stern-Gerlach setup,
we critically examine the relationship between the notion of
`distinguishability' of apparatus states defined in terms of the inner 
product and spatial separation among the emerging wave packets.
We show that, in general, the mutual orthogonality of these wave packets
does not necessarily imply their unambiguous spatial separation even
in the asymptotic limit. A testable scheme is formulated to quantify
such departures from `idealness' for a range of relevant parameters.
\end{abstract}


\maketitle

\section{Introduction}

Measurement theory in quantum mechanics deserves special attention compared to  that 
in classical mechanics. This is essentially due to the \emph{invasive character} of the measurement process 
since any measurement described by quantum mechanics necessarily entails an interaction between the observing apparatus 
and the observed system, and thus the state of the observed system is necessarily affected by the process of measurement. The quantum mechanical modelling for measurement process was first introduced 
by von Neuman\cite{von} where the measuring device was treated quantum mechanically. 

The essential theory of quantum measurement is as follows\cite{home}. Let the initial state of a system be given by
\begin{equation}
\phi_{0}=a\chi_{+}+b\chi_{-}
\end{equation}
where $\chi_{+}$ and $\chi_{-}$ are the mutually orthogonal eigenstates of a measured dynamical variable.
The initial combined state of the observed system and the apparatus is 
$\Psi_{i}=\phi_{0}\psi_{0}$ where $\psi_{0}$ is the apparatus state which 
is sharply peaked around the center of mass of position cordinates. After interaction with the measuring device, the final state is an entangled state which can be written as 
\begin{equation}
\Psi_{f}=a\psi_{+}\otimes\chi_{+}+b\psi_{-}\otimes\chi_{-}
\end{equation}     
where $\psi_{+}$ and $\psi_{-}$ are the apparatus states after interaction. Thus after measurement interaction,
there is one-to-one correpondence between the system and the apparatus 
states. 

Usually in any measurement situation the apparatus states are ultimately localized in position space. Now, for an `ideal measurement', it is required that the states $\psi_{+}$ and $\psi_{-}$
need to be \emph{macroscopically distinct} and \emph{mutually orthogonal} in the configuration space. Macroscopic distinguishability between apparatus states is a key notion in quantum measurement theory
which calls for careful scrutiny of its various subtleties in different 
experimental contexts\cite{leggett}.

In this context it is usually {\it assumed} that orthogonality between the states in configuration 
space {\it implies} distinguishability in the position space. Here in this paper we critically examine the above assumption in  the context of the Stern-Gerlach (SG) experiment \cite{gerlach}which is considered to be an archetypal 
example of quantum measurement. In particular, we study the operational compatibility between the orthogonality and position space  distinguishability of the appratus states. In the SG device employed for measuring 
the  spin of a quantum particle, the apparatus states are represented by
the spatial wave functions of the observed particles whose spins are inferred from the observed positions. As mentioned earlier, usually all experiments ultimately reduce to the macroscopic distinction of position (pointer reading, flash of light on a screen etc.). The SG experiment exibits \emph{perfect} correlation between two degrees of freedom of a single system in terms of position and spin so that the value of position \emph{definitely} 
allows us to infer the value of spin. 

The detection of quantized spin angular momentum of particles is not the only importance of the SG experiment. The SG interferometry has been an active area of research over past several decades. It has attracted attention of a number of well-known contributors like Bohm\cite{bohm}, followed by Wigner's work\cite{wigner} on the problem of reconstructing the initial state and its relevence to the issue of wave function collapse in quantum measurement. Subsequently, Englert, Schwinger  and Scully \cite{eng}  have analysed this issue in much depth (the well-known Humty-Dumpty problem) in a series of three papers. It has also been experimentally studied\cite{robert} how the extracted phase information from the SG interferometry experiment determines the transfer of coherence of spin to the external degree of freedom(position) giving rise to the position-spin entanglement. More investigation along this direction has been pursued by Oliviera and Calderia \cite{oliv} by using SQUID as the source of the magnetic field. Also, the usefulness of the SG experiment in probing more critically the subtleties of the relationship between which path detection and interference has been recently revisited\cite{reini} in the context of the works by Duerr\cite{duerr} and Knight\cite{knight}.

Our analysis reveals that contrary to the usual assumption, orthogonality in 
the configuration space does \emph{not always imply} distinguishability in the 
position space. Usually, it turns out that even with nonorthogonality in
the configuration space, the emergent wave packets are spatially well separated for
all practical purposes at sufficient distances. However, we show here
that there can be situations in which although $\psi_{+}$ 
and $\psi_{-}$ are orthogonal (a \emph{formally ideal} situation), but still 
there persists a finite overlap in position space between the associated 
wave packets of $\psi_{+}$ and $\psi_{-}$ even at asymptotic distances
(we call this as an 
\emph{operationally nonideal} situation). We use the results of the SG  
experimental setup in order to illustrate this issue of compatibility of
formal idealness with operational idealness. Moreover, we propose a scheme
to quantify the departure from the usual ideal situation that can be 
experimentally tested, and the magnitude of such a departure is measured  by suitably 
placing a subsequent ideal SG setup.  

The discussions regarding the nonideality of the SG experiment in the 
literature are mostly related to the practical problem involved in ensuring 
the required inhomogeneity of the magnetic field\cite{alstrom,cruz}. However,
the nonideality considered here arising due to finite
overlap between the emergent wave packets in position space not only illustrates 
a hitherto unexplored conceptual subtlety in the standard formalism of quantum mechanics, but can also, 
in practice, generate error in the inference of the value of spin
of a particle from its position  on the screen placed even at large distances. Thus, our study 
goes through a route which is completely different from the previous 
authors\cite{alstrom,cruz} who have
studied the issue of nonidealness in the context of \emph{how} to 
idealize the experiment. On the other hand, our present scheme is concerned with 
possible outcomes in the different types of nonideal
situations that we will describe in details. 

The plan of the paper is as follows. In section II, we present a 
general analysis of the measurement of the spin of spin-$1/2$ particles using the SG setup, 
which has not received enough in-depth attention in in the literature.
In this process we define precisely the above-mentioned 
notions of formal and operational idealness.
This sets the stage for formulating our scheme for
quantifying the departures from idealness  
which we study in section III. We identify two distinct categories of 
operational and formal nonidealness.
The actual computations of the measures of non-idealness
are done for a range of relevant experimental parameter values in section IV.
We provide illustrative numerical estimates showing differences of 
the outcomes between ideal and nonideal situations through
which we demonstrate the lack of universal
correspondence between orthogonality in configuration space and
distinguishability in position space in the SG experiment. Finally,  in 
Section V we give a summary of our results and concluding remarks. 

\section{Quantum measurement theory using the Stern-Gerlach device}

\begin{figure}[h]
{\rotatebox{0}{\resizebox{7.0cm}{4.0cm}{\includegraphics{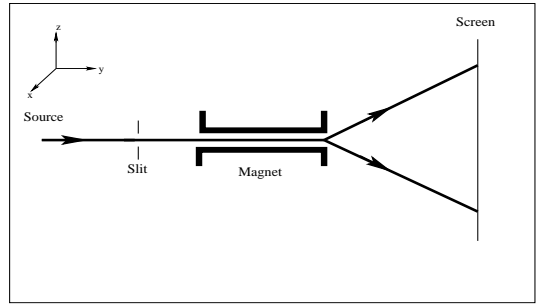}}}}
\caption{\label{Figure 1} {\footnotesize A schmetic Stern-Gerlach Setup}}
\end{figure}

The conventional account of the SG-experiment was given by Bohm\cite{bohm} 
and later others have studied the quantum theoretic treatment of the 
problem\cite{scully}.
The usual description of the \emph{ideal} SG experiment (Fig.1) is as follows.
A beam of \emph{x-polarized} spin-1/2 neutral particles, say neutrons, with 
finite magnetic moment is represented by the total wave function 
$\Psi\left(\textbf{x},t=0\right)=\psi_{0}\left(\textbf{x}\right)\chi(t=0)=\psi_{0}\left(\textbf{x}\right)\otimes\left(\alpha\left|\uparrow\rangle_{z}+\beta\right|\downarrow\rangle_{z}\right)$. The spin part $\chi(t=0)$ is the state of the system 
to be observed where $|\uparrow\rangle_{z}$ and $|\downarrow\rangle_{z}$ are 
the eigenstates of ${\sigma}_{z}$ and $\alpha$ and $\beta$ satisfy the 
relation $|\alpha|^{2}+|\beta|^{2}=1$. The spatial part $\psi_{0}({\bf x})$ 
represents the initial state of the measuring device and is associated with a 
Gaussian wave packet which is initially peaked at the entry point(${\bf x}=0$)
of the SG magnet and starts moving along $+ve$ $y-axis$ with velocity $v_{y}$ 
through a transversely directed (along $+ve$ $z-axis$) inhomogeneous magnetic 
field (localised between $y=0$ and $y=d$) with respect to the direction of 
the beam. Within the SG magnet, in addition to the $+\widehat y -axis$ motion 
the particles gain  velocity with magnitude $v_{z}$ along the 
$\widehat z-axis$  due to the interaction of their spins with the 
inhomogeneous magnetic field during the time $\tau$. 

The time evolved total wave function at the time $\tau$, which is an 
entangled state 
between position and spin is then given by 
$\Psi({\bf x},\tau)=\alpha\psi_{+}(\textbf{x},\tau)\otimes\left|\uparrow\right\rangle_{z}+\beta\psi_{-}(\textbf{x}, \tau)\otimes\left|\downarrow\right\rangle_{z}$. 
At the exit point ($y=d$) of the SG 
magnet, the particles deflect differently in a way that particles with 
eigenstate $\left|\uparrow\right\rangle_{z}$ associated with the wave packet 
$\psi_{+}({\bf x},\tau)$ move freely along the direction 
$\widehat n_{+}=v_{y}\widehat j+v_{z}\widehat k$ and the particles with 
eigenstate $\left|\downarrow\right\rangle$ associated with the wave packet 
$\psi_{-}({\bf x},\tau)$ move freely along the direction 
$\widehat n_{-}=v_{y}\widehat j-v_{z}\widehat k$ direction. 
Our first task is to evaluate the explicit expressions for 
$\psi_{+}({\bf x},t)$ and $\psi_{-}({\bf x},t)$ corresponding to 
$|\uparrow\rangle_{z}$ and $|\downarrow\rangle_{z}$ after emerging from the 
exit point ($y=d$) of the SG magnet. To this end we provide a fully quantum 
mechanical description of the theory of SG experiment here.  

We start our calculation by taking the initial total wave function of the 
particle at $t=0$ to be
\begin{equation}
\Psi\left(\textbf{x},t=0\right)=\psi_{0}\left(\textbf{x}\right)\chi(t=0)=\psi_{0}\left(\textbf{x}\right)\otimes\left(\alpha\left|\uparrow\rangle_{z}+\beta\right|\downarrow\rangle_{z}\right)
\label{init}
\end{equation}
where 
$\chi(t=0)=\alpha\left|\uparrow\rangle_{z}+\beta\right|\downarrow\rangle_{z}$ 
is the initial spin
state with $\left|\alpha\right|^{2}+\left|\beta\right|^{2}=1$ and 
$\left|\uparrow\right\rangle_{z} $,
$\left|\downarrow\right\rangle_{z}$ are the eigenstates of $\sigma_{z}$, 
and $\psi_{0}\left(\textbf{x}\right)$ is the initial spatial part of the 
total wave function represented by a Gaussian wave packet which is peaked at 
the entry point (${\bf x}=0$) of the SG magnet at $t=0$ given by 
\begin{equation}
\psi_{0}\left({\bf x}\right)=\frac{1}{{(2\pi\sigma_{0}^{2})}^{{3}/{4}}}
\exp\left(-
\frac{{\bf x}^{2}}{4\sigma_{0}^{2}}+ i{\bf k}.{\bf x}\right)
\label{initpack}
\end{equation}
where $\sigma_{0}$ is the initial width of the wave packet. The 
wave packet moves along +ve $y-axis$ with the initial group velocity $v_{y}$
and the wave number $k_{y}=\frac{m v_{y}}{\hbar}$.

The interaction Hamiltonian is $H_{int}=\mu\bm\sigma.\textbf{B}$ where $\mu$ 
is the magnetic moment of the neutron, $\textbf{B}$ is the inhomogeneous 
magnetic field and $\bm\sigma$ is the Pauli spin matrices vector. Then the 
time evolved total wave function  at $t=\tau$ after the interaction of 
spins with the SG magnetic field is given by
\begin{eqnarray}
\nonumber
\Psi\left(\textbf{x},t=\tau\right) &=& \exp({-\frac{iH\tau}{\hbar}})\Psi(\textbf{x},t=0)\\
&=&\alpha\psi_{+}(\textbf{x},\tau)\otimes\left|\uparrow\right\rangle_{z}+\beta\psi_{-}(\textbf{x}, \tau)\otimes\left|\downarrow\right\rangle_{z}
\label{timeevolved}
\end{eqnarray}
where $\psi_{+}\left({\bf x},\tau\right)$ and 
$\psi_{-}\left({\bf x},\tau\right)$
are the two components of the spinor 
$\psi=\left(\begin{array}{c}\begin{array}{c} 
\psi_{+}\\ \psi_{-}\end{array}\end{array}\right)$ which satisfies the 
Pauli equation. We take the inhomogeneous magnetic field as 
${\bf B}=(-bx,0,B_{0}+bz)$  satisfying the Maxwell equation 
${\bm \nabla}.{\bf B}=0$, instead of the field chosen in the original 
Stern-Gerlach paper\cite{gerlach} which was not divergence free. The 
two-component Pauli equatic§ can then be written as two  coupled
equations for $\psi_+$ abd $\psi_-$, given by 
\begin{eqnarray}
i\hbar\alpha\frac{\partial {\psi}_+ }{\partial t} = -\alpha\frac{\hbar^{2}}{2m}\bm\nabla^{2}{\psi}_{+}+\alpha\mu(B_0 + b z){\psi}_+ -\beta\mu b x {\psi}_- 
\nonumber\\
\label{coupledspinor}
\\
i\hbar\beta\frac{\partial {\psi}_- }{\partial t} = -\beta\frac{\hbar^{2}}{2m}\bm\nabla^{2}{\psi}_{-}+\alpha\mu b x {\psi}_+ -\beta\mu(B_0 + b z){\psi}_-
\nonumber
\end{eqnarray}

Due to the realistic magnetic field, there exists an equal force transverse 
to the $z-axis$, and a continuous distribution is expected instead of usual 
line distribution. But the time average of the transverse force along 
the $x-axis$
is zero due to the rapid precession of the magnetic moment around the field 
direction\cite{alstrom} provided that $B_0$ is much greater than 
the degree of inhomogeneity $b$. By using coherent internal states, 
it has been argued\cite{cruz} that the exact condition to neglect the 
tranverse 
component is $B_0\gg b\sigma_0$ where $\sigma_{0}$ is the width of the initial
wave packet. Using the above condition the coupling between the above two 
equations is removed and one obtains the following decoupled equations given by
\begin{eqnarray}
i\hbar\frac{\partial\psi_{+}}{\partial t}&=&-\frac{\hbar^{2}}{2m}\bm\nabla^{2}{\psi}_{+}+\mu\left(B_{0}+bz\right)\psi_{+} \nonumber\\
i\hbar\frac{\partial\psi_{-}}{\partial t}&=&-\frac{\hbar^{2}}{2m}\bm\nabla^{2}{\psi}_{-}-\mu\left(B_{0}+bz\right)\psi_{-}
\label{decoupled}
\end{eqnarray}
The solutions of the above equations can be written as
\begin{eqnarray}
\psi_{+}\left(\textbf{x};\tau\right)= \nonumber\\
\frac{1}{\left(2\pi s_{\tau}^{2}\right)^{\frac{3}{4}}}\exp\left[-\left\{ \frac{x^{2}+(y-v_{y}\tau)^{2}+(z-\frac{v_{z}\tau}{2})^{2}}{4\sigma_{0}s_{\tau}}\right\}\right]\nonumber\\
\times \exp\left[i\left\{-\Delta_{+}+ \left(y-\frac{v_{y}\tau}{2}\right)k_{y}+ k_{z}z\right\}\right]\nonumber\\
\psi_{-}\left(\textbf{x};\tau\right)= \nonumber\\
\frac{1}{\left(2\pi s_{\tau}^{2}\right)^{\frac{3}{4}}}\exp\left[-\left\{ \frac{x^{2}+(y-v_{y}\tau)^{2}+(z+\frac{v_{z}\tau}{2})^{2}}{4\sigma_{0}s_{\tau}}\right\}\right]\nonumber\\
\times \exp \left[i\left\{-\Delta_{-}+ \left(y-\frac{v_{y}\tau}{2}\right)k_{y}-k_{z}z\right\} \right]
\label{solutions}
\end{eqnarray}
where $\Delta_{\pm}=\pm\frac{\mu B_{0}\tau}{\hbar}+\frac{m^{2} v_{z}^{2}\tau^{2}}{6\hbar^{2}}$,~~$v_{z}=\frac{\mu b\tau}{m}$,~~ $k_{z}=\frac{m v_{z}}{\hbar}$ and ~~$s_{t}=\sigma_{0}\left(1+\frac{i\hbar t}{2m\sigma_{0}^{2}}\right)$.
Here $\psi_{+}\left(\textbf{x},\tau\right)$ and $\psi_{-}\left(\textbf{x},
\tau\right)$ representing the wave functions at the exit point ($y=d$) of the 
SG magnet at $t=\tau$
correspond to $\left|\uparrow\right\rangle_{z} $ and $\left|\downarrow\right\rangle_{z} $ respectively, with average momenta 
$\langle\widehat p\rangle_{\uparrow}$ and $\langle\widehat p\rangle_{\downarrow}$,
where $\langle\widehat p \rangle_{\uparrow\downarrow}=(0,mv_{y},\pm\mu b\tau)$.
Within the magnetic field the particles gain the same magnitude of momentum 
$\mu b\tau$ but the directions are such that the particles with eigenstates 
$|\uparrow\rangle_{z}$ and $|\downarrow\rangle_{z}$ get the drift along 
$+$ve $z$-axis and $-$ve $z$-axis respectively, while the $y$-axis momenta
remain unchanged. Hence after emerging from the SG magnet the particles
represented by the components
$\psi_{+}\left(\textbf{x},\tau\right)$ and $\psi_{-}\left(\textbf{x},\tau\right)$ move {\it freely}  along the respective directions  
$\widehat n_{+}=v_{y}\widehat j + \frac{\mu b\tau}{m}\widehat k$ and 
$\widehat n_{-}=v_{y}\widehat j - \frac{\mu b\tau}{m}\widehat k$ with 
the \emph{same} group velocity $v=\sqrt{v^{2}_{y}+(\frac{\mu b\tau}{m})^{2}}$ 
fixed by the parameters 
of the SG set-up and the initial velovity $(v_{y})$ of the peak of the 
wave packet. 

Now, the inner product $I$ between the $\psi_{+}({\bf x},\tau)$ and 
$\psi_{-}({\bf x},\tau)$ components is given by 
\begin{equation}
I=\int_{-\infty}^{+\infty}\psi^{*}_{+}({\bf x},\tau) \psi_{-}({\bf x},\tau)d^{3}{\bf x}
\label{innerprod}
\end{equation}
and is taken to be \emph{zero} for the \emph{formally ideal} situation. 
This inner product is preserved for the subsequent time evolution during which
the freely evolving wave functions at a time $t$ after emerging 
from SG setup are given by
\begin{eqnarray}
\psi_{+}({\bf x},t)=\frac{1}{(2 \pi s^{2}_{t+\tau})^{3/4}}\nonumber\\
\times \exp\left[-\left\{\frac{x^{2}+\left(y-v_{y}(\tau+t)\right)^{2}+\left(z-\frac{v_{z}\tau}{2}-v_{z}t\right)^2}{4 \sigma_0 s_{t+\tau}} \right\}\right]\nonumber\\
\times \exp\left[i\left\{-\Delta_{+} +k_{y}\left(y-\frac{v_{y}(\tau+t)}{2}\right)+k_{z}(z-\frac{v_{z}t}{2})\right\}\right] \nonumber\\
\psi_{-}({\bf x},t)=\frac{1}{(2\pi s^{2}_{t+\tau})^{3/4}}\\
\times \exp\left[-\left\{\frac{x^{2}+(y-v_{y}(\tau+t))^{2}+(z+\frac{v_{z}\tau}{2}+v_{z}t)^2}{4 \sigma_0 s_{t+\tau}} \right\}\right]\nonumber\\
\times \exp\left[i\left\{-\Delta_{-} +k_{y}\left(y-\frac{v_{z}(\tau+t)}{2}\right)-k_{z}\left(z+\frac{v_{z}t}{2}\right)\right\}\right]\nonumber
\label{freewavefn}
\end{eqnarray}
where $s_{t+\tau}=\sigma_{0}\left(1+\frac{i\hbar (t+\tau)}{2m\sigma_{0}^{2}}\right)$.
The free time evolved wave functions $\psi_{+}({\bf x},t)$ and 
$\psi_{-}({\bf x},t)$ after emerging from the SG magnet at time $t$ are 
\emph{orthogonal} in the \emph{formally ideal situation}. 

Let us now discuss the outcomes of this ideal situation from the formal and 
operational viewpoints. In a \emph{formally ideal} measurement $I=0$.  
After emerging from the exit point of the SG magnet
the probabilities of finding particles with up and down spin in the 
$z$-direction, i.e.,
$\left|\uparrow\right\rangle_{z}$ and $\left|\downarrow\right\rangle_{z}$,  
are $P_{\uparrow}^{i}=|\alpha|^{2}$ and  $P_{\downarrow}^{i}=|\beta|^{2}$ 
respectively. In order to discriminate the above situation from the case
of an \emph{operationally ideal} 
situation, we define \emph{operational idealness} by the condition that the 
probabilities of finding
particles within the $+$ve $zy$-plane (\emph{upper} plane) and $-$ve 
$zy$-plane (\emph{lower} plane) are $P_{+}^{i}=|\alpha|^{2}$ and  
$P_{-}^{i}=|\beta|^{2}$ respectively. 
Combining these two statements coming from the formal
and the operational viewpoints, we can say that when a measurement is 
\emph{both formally and operationally ideal} then $P_{+}^{i}=P_{\uparrow}^{i}$
and $P_{-}^{i}=P_{\downarrow}^{i}$, i.e., the probability of finding $\left|\uparrow\right\rangle_{z}$
particles equals the probability of finding particles in the upper plane,
and similarly for $\left|\downarrow\right\rangle_{z}$ particles and those in the lower plane.
In other words, in a perfectly (formally as well as operationally) ideal
Stern-Gerlach experiment, all $\left|\uparrow\right\rangle_{z}$ particles can be found in the 
upper plane, whereas all $\left|\downarrow\right\rangle_{z}$ particles can be found in the lower
plane. 

\section{The non-ideal Stern-Gerlach experiment}

In the context of the SG experiment the above discussed ideal situation is a 
very special case because, in general, \emph{orthogonality} between  
$\psi_{+}({\bf x},\tau)$ and $\psi_{-}({\bf x},\tau)$ crucially depends on 
the delicate choices of some relevant parameters involved in the SG setup. 
Substituting the expressions for 
$\psi_{+}\left(\textbf{x},\tau\right)$ and 
$\psi_{-}\left(\textbf{x},\tau\right)$ given by Eqs.(\ref{solutions}) in
Eq.(\ref{innerprod}), one obtains the actual expression for the 
inner product $|I|$ (the 
inner product may contain a global phase and hence we take the modulus of the 
inner product) to be 
\begin{equation}
|I|= exp\left\{-\frac{\mu^{2}b^{2}\tau^{4}}{8 m^{2}\sigma_{0}^{2}}-\frac{2\mu^{2}b^{2}\tau^{2}\sigma_{0}^{2}}{\hbar^{2}}\right\}
\label{innerprod2}
\end{equation}
which will be preserved after subsequent free time evolution. It is seen from 
Eq.(\ref{innerprod2}) that $|I|$ depends 
on the parameters $b$, $\tau$, $m$ and $\sigma_{0}$ and for 
{\it sufficiently large} values of $b$ and $\tau$ with fixed $\sigma_{0}$ and 
$m$, one has $|I| \approx 0$, i.e., $\psi_{+}\left({\bf x},t\right)$ and 
$\psi_{-} \left({\bf x},t\right)$ are {\it orthogonal} for all practical
purposes. But in 
general, as we will see in  the next section, there could be various choices 
of the relevant parametres for which $|I|\neq 0$.

Our purpose here is to explore the \emph{nonideal}  situation from the
viewpoints of both formal orthogonality and operational distinguishability
and investigate the connection between the two by
quantifying the departures from the ideal measurement outcomes. The question 
arises as to how one can predict the outcomes of this nonideal experiment. 
It is well-known that nonorthogonal states can \emph{not} be distinguished 
perfectly, even if they are known. There are various schemes\cite{ivanovic} 
for optimum discrimination among the states by adopting different strategies. 
Usually all experiments ultimately reduce to the 
measurement of position, and here in this work we are confined to the 
operational discrimination between 
the states in the position space.

From the operational viewpoint, the above question may be posed as folows: 
What is the probability of finding particles with 
$|\uparrow\rangle_{z}$ (or $|\downarrow\rangle_{z}$) in the \emph{lower plane}
(or \emph{upper plane}) when $|I|\neq 0$ ? In order to find an answer to this,
we {\it define} an error integral $E\left(t\right)$, the key ingredient in our
scheme which gives a quantitative prediction for this nonideal situation. 
The parameter $E(t)$ is a function of time and is given by 
\begin{eqnarray}
E(t)&=&\int_{x=-\infty}^{+\infty}\int_{y=-\infty}^{+\infty}\int_{z=0}^{+\infty}|\psi_{-}({\bf x},t)|^{2}dx dy dz\\
\nonumber
&&=\int_{x=-\infty}^{+\infty}\int_{y=-\infty}^{+\infty}\int_{z=-\infty}^{0}|\psi_{+}({\bf x},t)|^{2}dx dy dz
\end{eqnarray}
where $E(t)$ multiplied by $|\alpha|^{2}$ (or $|\beta|^{2}$) gives the 
probability of finding $|\downarrow\rangle_{z}$ (or $|\uparrow\rangle_{z}$ ) 
particles within the upper plane (or lower plane) at time $t$. It turns out 
from the solutions given by Eqs.(10) that the parameter 
$E(t)$ is \emph{not} zero just after the two wave packets emerge from the 
SG magnet at $t=\tau$, but during the course of free evolution $E(t)$ 
saturates to a minimum value, say $E_{s}$, with the saturation time $t_{s}$  
depending upon the choices of relevant parameters involved. It is then 
logical to consider $E_{s}$ as a measure of nonidealness. The value of 
$E_{s}$ varies between {\it zero} and {\it one-half}, depending 
upon the values of
the relevant parameters $b$, $\tau$, $m$ and $\sigma_0$,  so that $E_{s}=0$ 
represents the \emph{operationally ideal} situation, whereas $E_{s}=0.5$ the 
\emph{fully nonideal} one.

Note that $|I|$ is \emph{not} the measure of operational nonidealness but the
modified observable probability is concerned with the $E_{s}$. Now, 
the {\it modified observable probabilities} of finding
the particles with $\left|\uparrow\right\rangle_{z}$ (spin up) in the 
\emph{upper plane} and $\left|\downarrow\right\rangle_{z}$ (spin down) 
in the \emph{lower plane} under the nonideal situation are respectively 
given by
\begin{eqnarray}
&&P_{\uparrow}^{ni}=(1-E_{s})|\alpha|^{2}\nonumber\\
&&P_{\downarrow}^{ni}=(1-E_{s})|\beta|^{2}
\label{nonideal}
\end{eqnarray}
where $P_{\uparrow}^{ni}+P_{\downarrow}^{ni}\neq1$. In this case in the upper 
(or lower) plane we get a \emph{mixture} of particles with both spin states 
$|\uparrow\rangle_{z}$ and $|\downarrow\rangle_{z}$. Hence the 
probabilities of 
finding $|\downarrow\rangle_{z}$ particles in the upper plane and 
$|\uparrow\rangle_{z}$ particles in the lower plane are $E_{s}|\beta|^{2}$ and
$E_{s}|\alpha|^{2}$ respectively. Then the probabilities of finding both
$\left|\uparrow\right\rangle_{z}$ and $\left|\downarrow\right\rangle_{z}$ particles 
\emph{total probability in the upper plane} and the \emph{total probability in 
the lower plane} are respectively given by
\begin{eqnarray}
&&P_{+}^{ni}=(1-E_{s})|\alpha|^{2}+E_{s}|\beta|^{2}\nonumber\\
&&P_{-}^{ni}=(1-E_{s})|\beta|^{2}+E_{s}|\alpha|^{2}
\label{nonideal2}
\end{eqnarray}
where $P_{+}^{ni}+P_{-}^{ni}=1$ and $E_{s}=0$ gives the result of 
the ideal measurement.
These $P_{+}^{ni}$ and $P_{-}^{ni}$ constitute the basic observable 
probabilities in our scheme. To verify Eq.(\ref{nonideal2}) one needs to 
suitably place a subsequent \emph{usual ideal} SG setup with $|I|=0$ (at
a sufficiently large distance where the asymptotic condition $E_{s}=0$ is 
satisfied), which counts all particles in the upper plane. Then the 
probabilities of finding $\left|\uparrow\right\rangle_{z}$ and 
$\left|\downarrow\right\rangle_{z}$ are $(1-E_{s})|\alpha|^{2}$ and 
$E_{s}|\beta|^{2}$ respectively.

\vskip 0.2in

\begin{figure}[h]
{\rotatebox{0}{\resizebox{6.0cm}{4.0cm}{\includegraphics{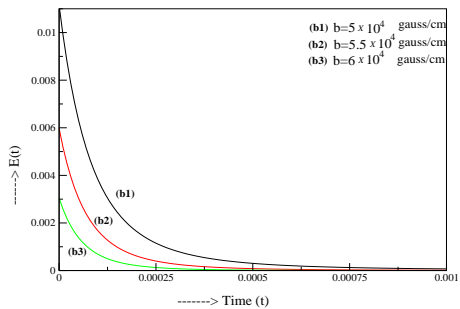}}}}
	\caption{\label{Figure 2}{\footnotesize The variation of $E(t)$ with 
time (in sec) is shown for three different values of $b$, i.e., 
$b=5\times10^4 gauss/cm$, $b=5.5\times10^4 gauss/cm$ and 
$b=6\times10^4 gauss/cm$ with $\tau=5\times10^{-4} sec$ and 
$\sigma_{0}=10^{-5}cm$ while $E_{s}=0$ and $|I|=0$. [CASE (i)]}}
\end{figure}

\vskip 0.1in 

As we have defined above,  $|I|=0$ implies the formally ideal situation, 
and $E_{s}=0$ 
the operationally ideal situation. Within the context of the nonideal 
Stern-Gerlach experiment, it is then possible to identify the 
following distinct situations which highlight the possible 
connections between configuration space orthogonality and position space
distinguishability.
 
(i) If the situation is \emph{operationally ideal} ($E_{s}=0$), then it 
\emph{must be} \emph{formally ideal} $(|I|=0)$. Or in other words, the 
observation of position
space distinguishability implies that the two wave functions are orthogonal
in configuration space. 

\vskip 0.2in

\begin{figure}[h]
{\rotatebox{0}{\resizebox{6.0cm}{4.0cm}{\includegraphics{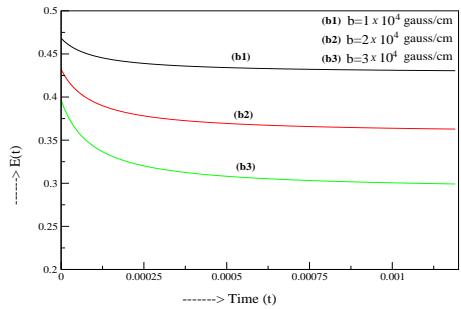}}}}
	\caption{\label{Figure 3}{\footnotesize The variation of $E(t)$ with 
time (in sec) is shown for the values $b=1\times10^3 gauss/cm$, 
$b=2\times10^3 gauss/cm$ and $b=3\times10^3 gauss/cm$, respectively, with 
$\tau=10^{-4} sec$ and $\sigma_{0}=10^{-4}cm$ while $E_{s}\neq0$ and 
$|I|\neq0$. [CASE (ii)]}}
\end{figure}

\vskip 0.1in

(ii) If the situation is \emph{formally nonideal} ($|I|\neq 0$), then it 
\emph{must be} \emph{operationally nonideal} ($E_{s}\neq0$). This means
that any non-orthogonality in the configuration 
space translates into the overlap
of the spatial wave functions, and this represents the  usual nonideal 
situation which has been studied by earlier authors\cite{alstrom,cruz} with 
the aim of
reducing the magnitude of nonidealness.

(iii) If the situation is \emph{formally ideal} ($|I|=0$),  it still 
\emph{may be}  \emph{operationally nonideal} ($E_{s}\neq0$). This is the most
interesting outcome of our present study since this \emph{hitherto unexplored}
nonideal situation implies that formal 
idealness or configuration space orthogonality does \emph{not always} 
guarantee operational idealness in terms of position space distinguishability.

\section{Quantitative estimates of operational versus formal nonidealness}

We will now show explicitly how the different situations (i), (ii) and (iii) 
arise due to the choices of the parameters in the SG experiment.
In order to illustrate these features, we present some numerical estimates for
the probabilities
$P_{\uparrow}^{ni}$ and $P_{\downarrow}^{ni}$, and $P_{+}^{ni}$ and 
$P_{-}^{ni}$ given in Eqs.(\ref{nonideal}) and (\ref{nonideal2}) respectively.
The estimation of these probabilities is contingent on the values of $\alpha$, 
$\beta$ and $E_{s}$. We first show three representative figures (Fig.2, Fig.3 
and Fig.4) corresponding to the situations (i), (ii) and (iii) respectively, 
which indicate how the  parameter $E(t)$ varies with time and saturates to 
$E_{s}$ (which is \emph{not} always zero). The curves in the figures are 
plotted by taking various choices of the relevant parameters, such as the 
degree of inhomogeneity of the  magnetic field $b$, and the interaction time 
$\tau$ while the initial width of the Gaussian wave packet $\sigma_{0}$ and 
the mass $m$ of the neutron are fixed.

\vskip 0.2in
 
\begin{figure}[h]
{\rotatebox{0}{\resizebox{6.0cm}{4.0cm}{\includegraphics{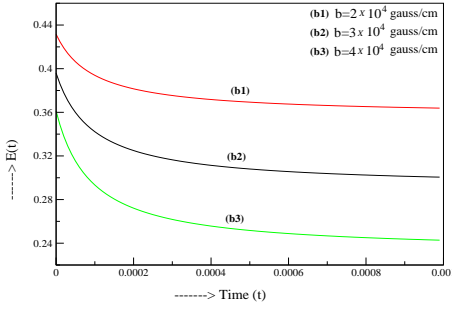}}}}
	\caption{\label{Figure 4}{\footnotesize The variation of $E(t)$ with 
time (in sec) is shown for the values $b=2\times10^4 gauss/cm$, 
$b=3\times10^4 gauss/cm$ and $b=4\times10^4 gauss/cm$, respectively, with 
$\tau=10^{-4} sec$ and $\sigma_{0}=10^{-5}cm$ while $E_{s}\neq0$ although 
$|I|=0$. [CASE (iii)]}}
\end{figure}

\vskip 0.1in

Corresponding to the above three cases, we
further plot the snapshots of the overlap between   
$|\psi_{+}(z,t)|^{2}$ and $|\psi_{-}(z,t)|^{2}$ (Figs.5, 6 and 7) for 
three different sets (Set-I, 
Set-II and Set-III) of the relevant parameters $\tau$, $b$, and $\sigma_{0}$ 
at two different times $t=10^{-5}sec$ and $t=0.1sec$. For Set-I, 
$b=6\times10^4 gauss/cm$, $\tau=5\times10^{-4} sec$ and $\sigma_{0}=10^{-5}cm$.
For Set-II, $b=2\times10^3 gauss/cm$, $\tau=10^{-4} sec$ and 
$\sigma_{0}=10^{-4}cm$ . For Set-III, $b=4\times10^4 gauss/cm$, 
$\tau=10^{-4}sec$ and $\sigma_{0}=10^{-5}cm$. These Set-I, Set-II and Set-III 
correspond to the three situations (i), (ii) and (iii) respectively as 
discussed earlier. One can see from Fig.7 that there exists a finite 
and appreciable overlap between the $|\psi_{+}({\bf x},t)|^{2}$ and 
$|\psi_{-}({\bf x},t)|^{2}$ at $t=10^{-5}sec$ which does \emph{not always} 
vanish at $t=0.1 sec$ (which is much larger than the saturation time $t_{s}$),
although the inner product $|I|$ is \emph{zero}.

\vskip 0.2in

\begin{figure}[h]
{\rotatebox{0}{\resizebox{7.0cm}{3.5cm}{\includegraphics{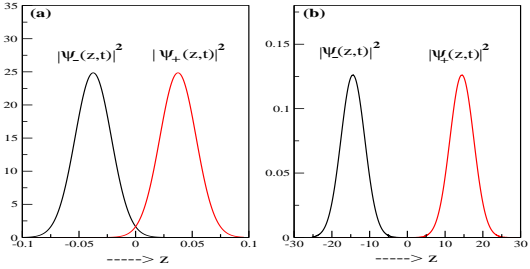}}}}
\caption{\label{Figure 5} {\footnotesize The overlap between   
$|\psi_{+}(z,t)|^{2}$ and $|\psi_{-}(z,t)|^{2}$ is plotted for 
$b=6\times10^4 gauss/cm$, $\tau=5\times10^{-4}sec$ and $\sigma_{0}=10^{-5}cm$ 
at two different times (a) $t=10^{-5}sec$, and (b) $t=0.1sec$ while $|I|=0$ 
and $E_{s}=0$. [CASE (i)]}}
\end{figure}

\vskip 0.1in

We now use the parameters of Set-III to calculate the probabilities for 
finding spin up $P_{\uparrow}^{ni}$ and spin down  
$P_{\downarrow}^{ni}$ particles from Eq.(\ref{nonideal}) in the formally 
ideal but 
operationally nonideal situation [CASE (iii)] and the corresponding 
probabilities $P_{+}^{ni}$ and $P_{-}^{ni}$ in the upper and lower planes,
respectively from Eq.(\ref{nonideal2}). We choose
four different values for $\alpha$ and $\beta$ satisfying 
$|\alpha|^2+|\beta|^2=1$. The saturation value of $E_{t}$ is obtained to be 
$E_{s}=0.2478$. The results are presented in Table 1.

\vskip 0.2in

\begin{figure}[h]
{\rotatebox{0}{\resizebox{7.0cm}{3.5cm}{\includegraphics{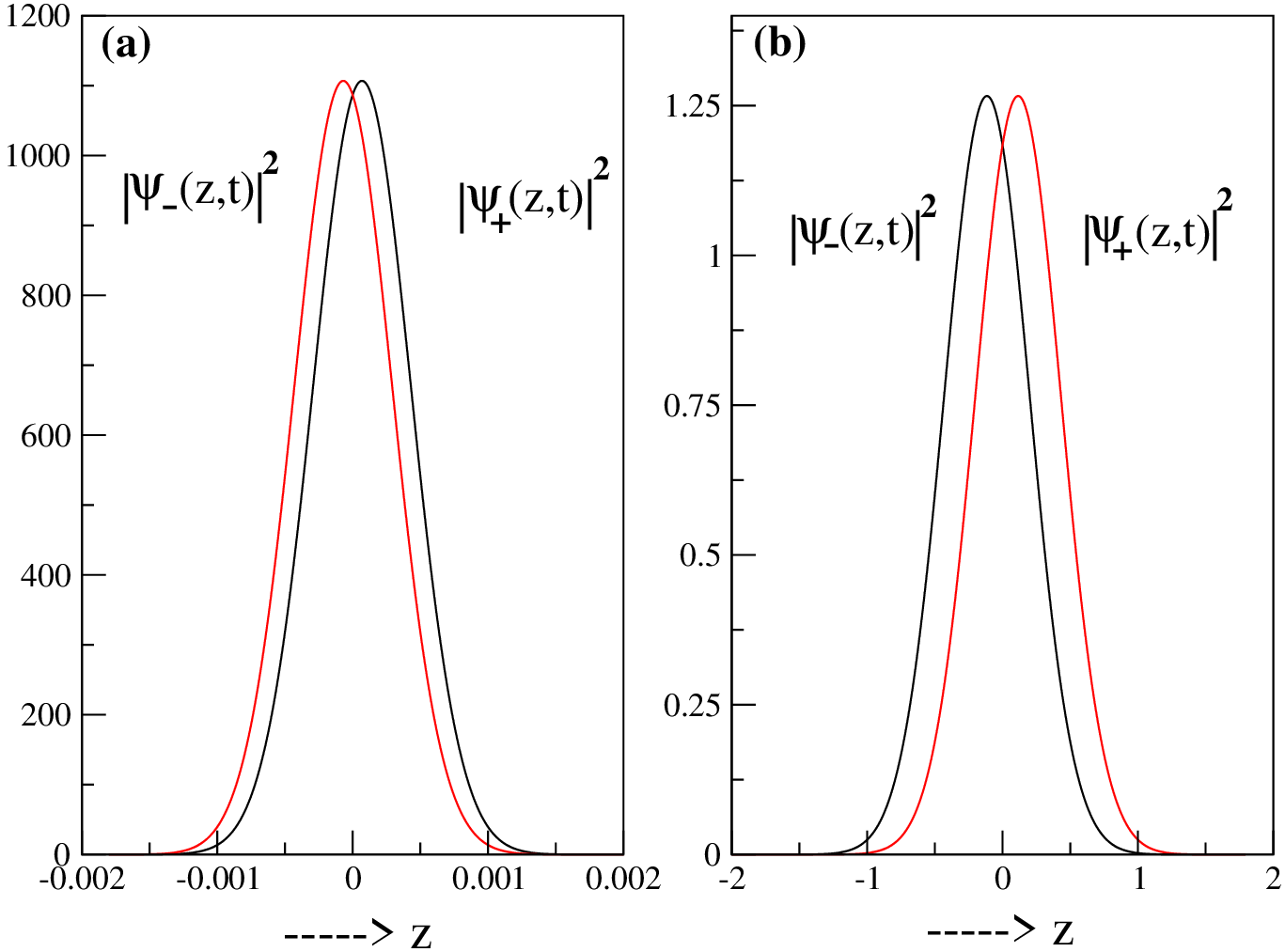}}}}
\caption{\label{Figure 6} {\footnotesize The overlap between 
$|\psi_{+}(z,t)|^{2}$ and $|\psi_{-}(z,t)|^{2}$ is plotted for 
$b=2\times10^3 gauss/cm$, $\tau=10^{-4} sec$ and $\sigma_{0}=10^{-4}cm$ 
at two different times (a) $t=10^{-5}sec$, and (b) $t=0.1sec$ while 
$|I|\neq0$ and $E_{s}\neq0$. [CASE (ii)]}}
\end{figure}

\vskip 0.1in

It is seen from the Table~1 that when $\alpha=\beta=1/\sqrt{2}$ the 
probability of finding particles with \emph{both} 
$\left|\uparrow\right\rangle$ and $\left|\downarrow\right\rangle$ spins
\emph{in the upper plane} is $P_{+}^{ni}=0.5000$ where probability of finding 
$\left|\uparrow\right\rangle$ (and $\left|\downarrow\right\rangle$) 
\emph{in the upper plane} is $P_{\uparrow}^{ni}=0.3761$ (and 
$=0.1239$). Note that, in the usual ideal situation (i) the
probability of finding particles with $\left|\uparrow\right\rangle$ in the 
upper plane is $P_{+}^{i}=P_{\uparrow}^{i}=0.5000$. To test the result 
experimentally, a subsequent SG setup which is \emph{ideal} in the sense of 
our situation (i), i.e., $|I|=0$ and $E_{s}=0$,  needs to be 
suitably placed. The
position of the second SG setup as well as the final screen position must be 
beyond the  corresponding saturation position $Y_{s}=v_{y} t_{s}$. The vaule 
of $t_{s}$ and subsequently $E_{s}$ is different for different parameter 
choices. For the parameters chosen in the Table~1, $t_{s}=0.0012 sec$ and 
hence the possible position of the second SG setup $Y_{s}$ is beyond 
$12 cm$ if one takes $v_{y}=10^{4}cm/sec$. 

\vskip 0.2in

\begin{figure}[h]
{\rotatebox{0}{\resizebox{7.0cm}{3.5cm}{\includegraphics{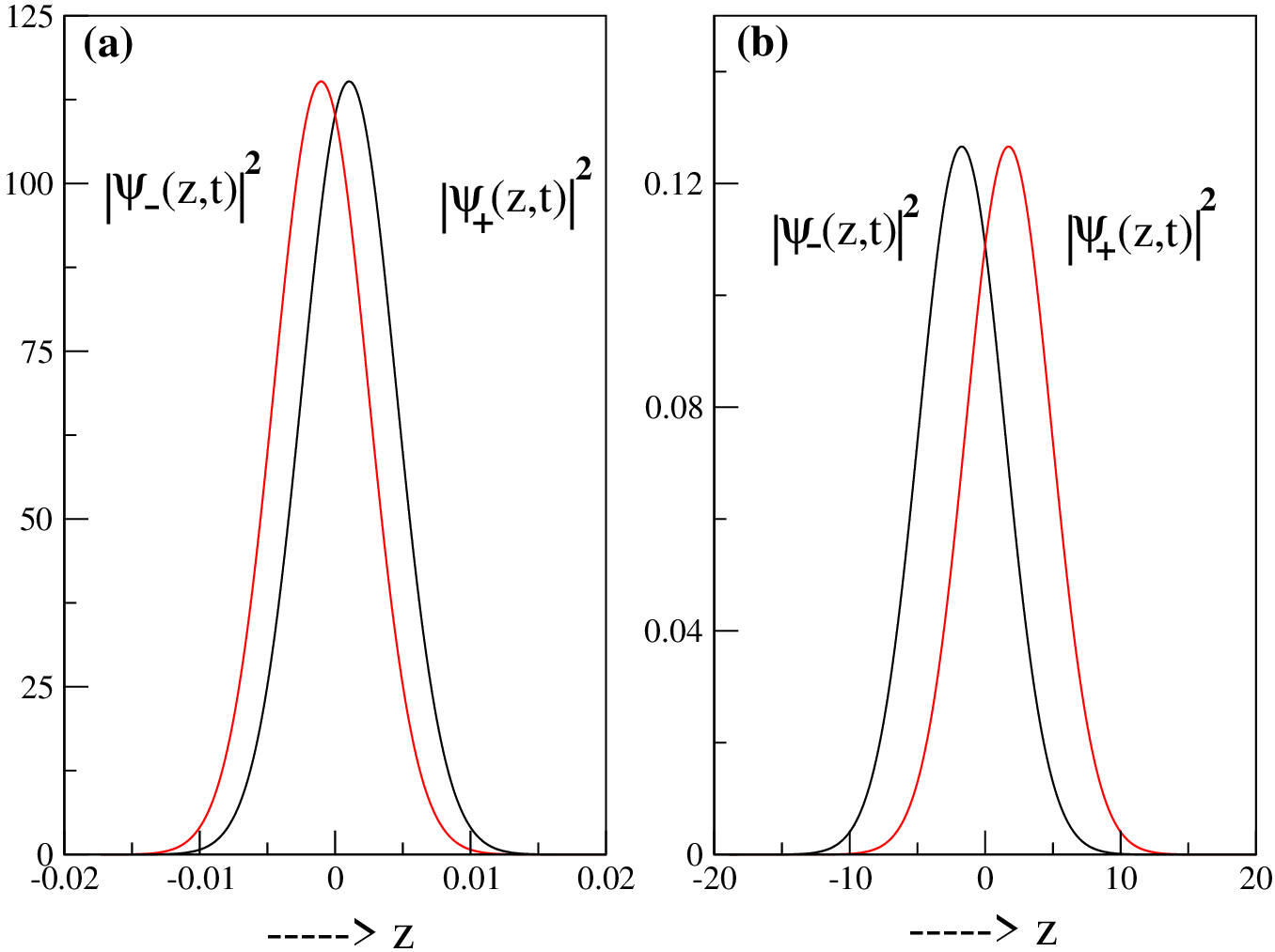}}}}
\caption{\label{Figure 7} {\footnotesize The overlap between the  
$|\psi_{+}(z,t)|^{2}$ and $|\psi_{-}(z,t)|^{2}$ is plotted for 
$b=4\times10^4 gauss/cm$, $\tau=10^{-4}sec$ and $\sigma_{0}=10^{-5}cm$ 
at two different times (a) $t=10^{-5}sec$, and  (b) $t=0.1sec$ while 
$|I|=0$ but $E_{s}\neq0$. [CASE (iii)]}}
\end{figure}

\vskip 0.1in

\begin{longtable}{|c|c|c|c|cc|cc|}
\hline 
$\alpha$&
$\beta$&
$P_{\uparrow}^{i}=P_{+}^{i}$&
$P_{\downarrow}^{i}=P_{-}^{i}$&
$P_{\uparrow}^{ni}$&
$P_{\downarrow}^{ni}$&
$P_{+}^{ni}$&
$P_{-}^{ni}$\tabularnewline
\hline
\hline 
$1/\sqrt{2}$&
$1/\sqrt{2}$&
0.5000&
0.5000&
0.3761&
0.3761&
0.5000&
0.5000\tabularnewline
0.8000&
0.6000&
0.6400&
0.3600&
0.4814&
0.2708&
0.5706&
0.4294\tabularnewline
$\sqrt{3}/2$&
$1/2$&
0.7500&
0.2500&
0.5642&
0.1881&
0.6261&
0.3739\tabularnewline
0.9487&
0.3162&
0.9000&
0.1000&
0.6770&
0.0752&
0.7018&
0.2982\tabularnewline
\hline
\end{longtable}
\footnotesize{TABLE.1: The quantities $P^i_{\uparrow}$ and $P^i_{\downarrow}$ 
denote the observable probabilities for finding respectively, the spin up and 
spin down particles corresponding to 
$\left|\uparrow\right\rangle_{z}$ and $\left|\downarrow\right\rangle_{z}$ 
of the \emph{ideal} SG measurement and $P^{ni}_{\uparrow}$ and 
$P^{ni}_{\downarrow}$ are the same for the \emph{nonideal} case. 
The quantities $P^i_{+}$ and $P^i_{-}$ denote the observable probabilities 
of finding particles \emph{in the upper plane} and \emph{in the lower plane} 
of the \emph{ideal} SG measurement and $P^{ni}_{+}$ and $P^{ni}_{-}$ are the 
same for the \emph{nonideal} case. Note that $P^i_{\uparrow}=P^i_{+}$ and 
$P^i_{\downarrow}=P^i_{-}$. In this Table the results are presented for four different choices of $\alpha$ and $\beta$ satisfying $|\alpha|^2+|\beta|^2=1$ 
with $E_{s}=0.2478$ for the relevant parameters $b=4\times10^4 gauss/cm$, 
$\tau=10^{-4} sec$ and $\sigma_{0}=10^{-5}cm$ while $|I|=0$. [CASE (iii)]}


\normalsize{

\section{Summary and Conclusions}

In this paper we have probed the usually implied assumption in the theory of quantum measurement that the configuration space orthogonality between the wave functions necessarily entails their unambiguous position space separability. While the latter is a key operational concept used to infer the outcomes of the SG experiment, our results show that the validity of the above assumption can be ensured only for specific choices of relevant parameters. It is demonstrated that there is indeed a range of possible choices that can lead to non-ideal situations where the above assumption is clearly falsified. Thus even a \emph{formally ideal} measurement situation in quantum mechanics
where the states are orthogonal in configuration space \emph{does not necessarily imply} an
unambiguous separation of  states in the position space which characterises an \emph{operationally ideal} measurement.

The non-idealness of the above kind where the overlap between the wave packets
in the position space persists even at large distance in spite of the 
orthogonality of the corresponding
configuration space states has remained hitherto unexplored.
It is thus important to evaluate theoretically in a formally 
ideal $|I|= 0$ but operationally non-ideal$(E_{s}\neq0)$ situation as to what will be the possible outcomes of the 
SG experiment when it is used in its various applications  \cite{bohm,wigner,eng,robert,oliv,reini,duerr,knight} mentioned in the beginning.
For this purpose we've defined an error integral $E(t)$ which is used to 
quantify the  results of the nonideal situations where the probabilities for
finding spin up particles in the lower plane and for finding spin down particles in the
upper plane are both non-vanishing. We find that the saturation value of
the error integral $E_{s}$ can be non-zero for a wide range of
the relevant parameters such as the inhomogeneity of the magnetic field $b$
and the SG interaction time $\tau$. Thus, in non-ideal situations, we can predict the observable
outcomes, i.e., the probabilities $P_{+}^{ni}$ and $P_{-}^{ni}$ corresponding
to both spin up and spin down particles in the upper and lower planes respectively.  These predictions can be experimentally verified by
placing a subsequent completly ideal (both formally and operationally) 
SG device at a suitable distance in accordance with the values of the relevant parameters.

The utility of such a scheme for quantifying the nonidealness in any given
SG setup lies in enabling the estimation of error involved in inferring the
measured spin state (i.e., the error in the state reconstruction process) 
from the actual measurement results. That this situation
may arise even in the formally ideal case with orthogonal states adds 
further practical relevance to the estimation of the observable outcomes.

To summarize, we have demonstrated that in a nonideal SG setup, 
conditions can be achieved for suitable choices of relevant parameters so 
that a significant spatial overlap between the emerging wave packets persists 
even when the inner product between the emergent wave functions is zero. 
Potential uses of variants of this type of \emph{nonideal } SG setup as a resource need to be 
explored in detail, particularly in regard to foundational issues such as 
the contentious question as to when a quantum measurement is \emph{completed}\cite{home}.
Furthermore, the analysis of operational non-idealness is likely to have quantitative implications in the SG interferometry. For example, one can attempt a non-ideal variant of an interesting example of quantum-state reconstruction \cite{hradil} involving analysis of a synthesis of noncommuting observables of spin-1/2 particles using SG device with varying orientations. As indicated by Hradil \emph{et al.} \cite{hradil}, the formalism developed for the analysis of such examples can be applied to the study of various problems like the estimation of the quantum state inside split beam neutron interferometers. Finally, we note that since the effect of enviournment induced decoherence on the position space overlap of the wave packets has been studied in an ideal SG setup\cite{venu}, it should be interesting to investigate such effects in the types of non-ideal SG setup discussed in this paper.

{\bf Acknowledgements}

We are grateful to John Corbett for helpful discussions. DH acknowledges 
 support from the Jawaharlal Nehru Fund, India. AKP acknowledges 
 support from the Council of Scientific and Industrial Research, India.

\end{document}